\documentclass[reprint,nofootinbib,amsmath,amssymb,aps,prl]{revtex4-1}

\usepackage{graphicx}
\usepackage{dcolumn}
\usepackage{bm}

\begin{document}

\title{Electrodynamic effect of anisotropic expansions in the Universe}

\author{Paolo Ciarcelluti}
 \email{paolo.ciarcelluti@gmail.com}
\affiliation{%
 Web Institute of Physics, www.wiph.org\\
 \homepage{http://www.WIPh.org}
}%
\affiliation{
 IFPA, D\'ep. AGO, Universit\'e de Li\`ege, 4000 Li\`ege, Belgium
}%

\date{\today}

\begin{abstract}
In the presence of anisotropic cosmic expansions at global or local scale the equations of electrodynamics in expanding space-time are modified and presented here.
A new effect should arise in regions of local anisotropic expansion in a cosmologically isotropic background. 
These regions should naturally exist, being connected with scales decoupling from the Hubble flow. 
Possible observational consequences of this effect are suggested.
In particular, I predict the appearance or variation of the polarization of electromagnetic radiation coming from or passing through these regions.
This effect is observable and possibly already observed in the polarization of quasars.
\end{abstract}

\pacs{Valid PACS appear here}
\maketitle

The Universe as a whole is usually considered as homogeneous and isotropic.
In fact the picture of its surface of last scattering, furnished at early times by the Cosmic Microwave Background (CMB), shows a very smooth distribution of matter, with inhomogeneities of order $10^{-5}$.
By contrast, at the present epoch the cosmic distribution of matter is very far from homogeneity, with a richness of structures at scales below a few hundreds Mpc, at which we observe clusters of galaxies, filaments, voids \cite{2002coec.book.....C}.
Observations show that the Cosmological Principle of isotropy and homogeneity is globally but not locally valid in the Universe.

Even with this evident inhomogeneity, on sufficiently large scales we observe a broadly isotropic expansion of the Universe, known as Hubble flow.
Despite this, there are several works that explore the possibility of cosmic anisotropies on the largest observable scales, connected with a number of observations in different astrophysical contexts.

There are several anomalies in the CMB maps, an extremely cold spot \cite{Naselsky:2007yd}, unusual alignments of the largest harmonic modes (sometimes called ``axis of evil'') 
\cite{Land:2006bn}, and a global hemispherical power asymmetry \cite{Kim:2010gf} (for a review, see for example Ref.\cite{Copi:2010na}; for a different statistical analysis see Ref.\cite{Samal:2008nv}).
All of them may represent evidence of cosmological anisotropy on large scales.

The polarization vectors of optical light emitted by distant quasars are not randomly oriented over the sky, but appear to be coherently aligned over huge regions of Gpc scale 
(``Hutsemekers effect'') 
\cite{Hutsemekers:1998,Hutsemekers:2000fv,Hutsemekers:2005iz,Jain:2003sg}.
In addition, the preferred direction shows a quasi-periodic dependence on the distance of the source, which means that the rotation of the polarization is linear with that distance \footnote{
The observation of a similar effect at radio wavelengths has been controversial\cite{Joshi:2007yf}, even if Ref.\cite{Nodland:1997cc} showed an indication of anisotropy based on a different study of the polarized radiation emitted by distant radio quasars.
The possible explanation can reside in different processes arising for optical and radio wavelengths during the propagation, which could randomize the polarization vectors and oppose to their alignment.
However, a very recent analysis based on the JVAS/CLASS radio surveys\cite{Tiwari:2012rr} shows a clear evidence of the same polarization effect observed at optical wavelengths.}.
It is interesting to note that the preferred axis for the alignment seems to be near the preferred directions suggested from CMB maps.

Using the observed motions of galaxies measured by the HST Extragalactic Distance Scale Key Project, the authors of Refs.\cite{2004mmu..sympE...5M,McClure:2007vv} found the existence of a statistically significant variation in the Hubble expansion rates across the sky, with differences between within and beyond our supercluster.

Luminosity distance measurements of high redshift SuperNovae Ia (SNIa) are a powerful tool to map the expansion flow at large scale.
Analyses of SNIa observations of High-z Supernova data \cite{Bochner:2004zg} and Union2 dataset \cite{Antoniou:2010gw,Cai:2011xs} show mild evidence of anisotropy. 
When combined with the preferred directions of other cosmological observations, the statistical evidence becomes stronger \cite{Antoniou:2010gw}.

There are essentially two possibilities to realize anisotropy in the Universe, either globally or ``locally''\footnote{
Here ``local'' is intended on cosmological scales, and for us usually refers to scales of order hundred Mpc.} 
(or both), according to the different sources of anisotropy.

On the global scale there are well studied homogeneous anisotropic cosmological models, the solutions of Einstein equations commonly known as Bianchi models \cite{2002coec.book.....C}.
Several papers, also recently \cite{Pontzen:2007ii,Sung:2010ek}, have tested their compatibility and tried to constrain their parameters, using CMB observations.
An alternative way to create a global anisotropic Hubble expansion is related to the existence of anisotropic dark energy (see, for example, Ref.\cite{Campanelli:2011uc}), which would drive expansion rates dependent on the direction, and create an ellipsoidal Universe.

A less explored possible anisotropy is at local scales.
In a homogeneous Universe there is a global expansion, but since the mass is non-uniformly distributed, also the expansion will be non-uniform, with different expansion rates at different places.
Thus the local anisotropic expansion is caused by inhomogeneities decoupling from the global expansion.
The gravitational attraction of the mass concentration in a supercluster of galaxies is expected to slow down the expansion around it.
A cluster or other structure embedded in the Hubble flow at some point starts to decelerate from the general expansion, and then eventually reaches turnaround and collapses in a bound structure decoupled from the global flow.
On the contrary, the expansion in underdense regions (voids) is less decelerated than in overdense regions, and then appears as accelerated with respect to the Hubble flow.
This effect has been invoked to try to explain the apparent accelerated expansion of the Universe \cite{Zehavi:1998gz,Bene:2003fz,Kolb:2004am}.
Another way to realize local anisotropies is to hypothesize the existence of fluctuations in dark energy density (see, for example, Ref.\cite{Urban:2010wa}).
These fluctuations would cause different accelerated expansions in different regions with the same scales of the dark energy density variations.

Whatever the anisotropies are, here I show their effect on the laws of electrodynamics.

I use Greek indices $\mu, \nu, ...$ for space-time coordinates, and Roman indices $i,j,...$ for pure space coordinates. 
Repeated indices are summed over all the coordinates.
I adopt units where the speed of light $c=1$ and a metric signature $(-,+,+,+)$.
$F_{\mu\nu}$ is the electromagnetic (EM) field tensor and $J^\mu \equiv (\rho_q,j^1,j^2,j^3)$ is the four-current density.

In an anisotropic expanding space-time the metric ${\rm d}s^2 = g_{\mu \nu} {\rm d}x^\mu {\rm d}x^\nu$ can be expressed locally as 
\begin{equation}
{\rm d}s^2 = -{\rm d}t^2 + a_1^2(t) ({\rm d}x^1)^2 + a_2^2(t) ({\rm d}x^2)^2 + a_3^2(t) ({\rm d}x^3)^2 ,
\label{metric}
\end{equation}
where $t$ is the proper time as measured by any fundamental observer, $x^1$, $x^2$ and $x^3$ are the comoving spatial coordinates, $a_1(t)$, $a_2(t)$ and $a_3(t)$ the scale factors in the three spatial directions.
The four-velocity corresponding to a fundamental observer in comoving coordinates is $u^\alpha \equiv (1,0,0,0)$.
The local expansion of the Universe is described by the three Hubble rates $H_i(t) = {\dot a_i}/{a_i}$.
The mean expansion rate (that defines the Hubble flow) is:
\begin{equation}
\bar{H} = \frac{1}{3} \sum_i \frac{\dot a_i}{a_i} = \frac{\dot{\bar{a}}}{\bar{a}}\;,
\end{equation}
where $\bar a = (a_1\,a_2\,a_3)^{1/3}$ is the geometric average of the expansion rates.

Following Ref.\cite{Subramanian:2009fu}, the EM tensor and its dual can be written in terms of the electric and magnetic fields as
\begin{eqnarray}
F_{\mu\nu} &=&  u_\mu E_\nu - u_\nu E_\mu
+ \epsilon_{\mu\nu\alpha\beta}B^\alpha u^\beta \;,
\label{fem}
\\
^*F^{\alpha\beta} &=& \frac12 \epsilon^{\alpha\beta\mu\nu}F_{\mu\nu}
\nonumber \\
&=&
\epsilon^{\alpha\beta\mu\nu}
u_\mu E_\nu
+ (u^\alpha B^\beta - B^\alpha u^\beta ) \; ,
\label{dualfem}
\end{eqnarray}
where $\epsilon^{\alpha\beta\mu\nu}$ is the totally antisymmetric
Levi-Civita tensor,
\begin{equation}
\epsilon^{\alpha\beta\mu\nu}=\frac{1}{\sqrt{-g}}
{\cal A}^{\alpha\beta\mu\nu} \; ,
 \quad
\epsilon_{\mu \nu \rho \lambda }=\sqrt{-g}
{\cal A}_{\mu \nu \rho \lambda },
\end{equation}
$g$ is the determinant of the metric tensor, and ${\cal A}^{\alpha\beta\mu\nu}$ is
the totally antisymmetric symbol such that ${\cal A}^{0123}=1$ and $\pm 1$ for
any even or odd permutations of $(0,1,2,3)$ respectively (note that ${\cal A}_{0123}=-1$).
Using our anisotropic metric (\ref{metric}), the non-vanishing independent components of the EM field tensor, which is antisymmetric and thus has vanishing diagonal components ($F^{\mu\mu}=0)$, are:
\begin{eqnarray}
F^{0i}&=&-F^{i0}=E^i, 
\nonumber \\
F^{12}&=&-F^{21}=\frac{a_3}{a_1\,a_2}\,B^3 \;,
\nonumber \\
F^{13}&=&-F^{31}=-\frac{a_2}{a_1\,a_3}\,B^2 \;, 
\nonumber \\
F^{23}&=&-F^{32}=\frac{a_1}{a_2\,a_3}\,B^1  \;.
\end{eqnarray}

\noindent Using the tensorial formulation of the Maxwell equations,
\begin{eqnarray}
F^{\mu \nu}_{\quad;\nu} &=& 4\pi J^\mu
\label{max1}
\nonumber \\
^*F^{\mu\nu}_{\quad ;\nu} &=& 0 \;,
\label{max2}
\end{eqnarray}
one derives the equations of electrodynamics in an anisotropic expanding space-time
\footnote{
An equivalent form of these equations was independently derived by Yves De Rop (private communication).}:
\begin{eqnarray}
&&\frac{\partial E^i}{\partial x^i} = 4\pi \rho_q \;,
\qquad 
  \frac{\partial B^i}{\partial x^i} = 0 \;,
\nonumber \\
&&\frac{1}{\bar a^3} \frac{\partial \left(\bar a^3 E^i\right)}{\partial t}
  = \frac{1}{\bar a} \epsilon_{ilm} \left(\frac{a_m}{\bar a}\right)^2 \frac{\partial B^m}{\partial x^l} - 4 \pi j^i \;,
\nonumber \\
&&\frac{1}{\bar a^3} \frac{\partial \left(\bar a^3 B^i\right)}{\partial t}
  = -\frac{1}{\bar a} \epsilon_{ilm} \left(\frac{a_m}{\bar a}\right)^2 \frac{\partial E^m}{\partial x^l} \;.
\label{maxanis}
\end{eqnarray}
Here $\epsilon_{ijk}$ is defined the 3-d fully antisymmetric symbol with $\epsilon_{123} = 1$.
For the derivation of the above equations I used the expressions of the connections in our metric (\ref{metric}) together with the antisymmetry of the EM tensor and its dual, so that:
\begin{eqnarray}
&&\Gamma^i_{\nu\alpha}\,F^{\alpha\nu} = \Gamma^i_{\nu\alpha}\,^*F^{\alpha\nu} = 0 \;,
\nonumber \\
&&\Gamma^0_{\nu\alpha}\,F^{\alpha\nu} = \Gamma^0_{\nu\alpha}\,^*F^{\alpha\nu} = 0 \;,
\nonumber \\
&&\Gamma^\nu_{\nu\alpha} = \frac{1}{2} g^{ii} \partial_\alpha g_{ii} = \Gamma^i_{i0} 
                       = \sum_i \frac{\dot a_i}{a_i} = 3\,\frac{\dot{\bar{a}}}{\bar{a}} = \Theta \;.
\end{eqnarray}
$\Theta$ is usually referred to as the expansion scalar and measures the fractional rate at which the volume 
changes with time for a central comoving observer.
Comparing Eqs.(\ref{maxanis}) with the corresponding equations (33) obtained in Ref.\cite{Subramanian:2009fu} for isotropic expanding space-time, the similarity is evident.
The differences are the obvious replacement of $a(t)$ with $\bar a(t)$, and the presence of the additional factor $\left({a_m}/{\bar a}\right)^2$ in the curl terms.
This factor is the scale factor in one direction relative to the average scale factor, and expresses the relative expansion in one direction.
It introduces a weight in the terms containing curls which is dependent on the direction of the field component.
These mixed scaling terms create an additional dependence on the directions of expansions that connect in a non-trivial way the equations of the three space components.

These equations take a more useful and transparent form in the natural reference frame for astronomical observations, consisting of the Fermi normal coordinates \cite{1963JMP.....4..735M}. 
They form the local inertial frame identified by the set of orthonormal tetrads tangent to the world line of the observer, and have the interesting property that the connections $\Gamma^\sigma_{\mu\nu}$ vanish at each point along the geodesic \cite{1973grav.book.....M}. 
In this frame the effect of the cosmological expansion is negligible locally and grows with distance.
In Fermi normal coordinates the time is the proper time $t$, the space coordinates are the proper space coordinates $r^i=a_i(t)\,x^i$, and the EM fields and the currents are represented as their projections along the orthonormal tetrads, that are simply given by
\begin{equation}
\bar{E}^i = a_i(t)\,E^i \;,\;\;\;\;\bar{B}^i = a_i(t)\,B^i \;,\;\;\;\;\bar{j}^i = a_i(t)\,j^i \;.
\end{equation}
Thus the Maxwell equations in an anisotropic expanding space-time become:
\begin{eqnarray}
&& \frac{\partial \bar{E}^i}{\partial r^i} = 4\pi \rho_q\;,
\qquad 
\frac{\partial \bar{B}^i}{\partial r^i} = 0 \;,
\nonumber \\
&&
\frac{\partial \left(\alpha_i^2 \bar{E}^i\right)}{\partial t}
= \epsilon_{ilm} \left(\frac{\alpha_i}{\alpha_m}\right)^2 \frac{\partial ({\alpha_m}^2 \bar{B}^m)}{\partial r^l} 
- 4 \pi \bar{j}^i \alpha_i^2 \;,
\nonumber \\
&&
\frac{\partial \left(\alpha_i^2 \bar{B}^i\right)}{\partial t}
= -\epsilon_{ilm} \left(\frac{\alpha_i}{\alpha_m}\right)^2 \frac{\partial ({\alpha_m}^2 \bar{E}^m)}{\partial r^l} \;,
\label{maxanis2}
\end{eqnarray}
where we have defined $\alpha_i^2 = \bar a^3 / a_i = a_l \, a_m$, which represents the geometric average of the two scale factors in directions perpendicular to the considered one.
The right hand sides of the last two equations are not simplified in order to make their meaning more explicit.
The similarity with the isotropic case, as expressed for example in Eqs.(37) of Ref.\cite{Subramanian:2009fu}, is easy to see if we replace $a(t)$ with the scale factor in the corresponding direction $a_i(t)$ and if we forget for a moment the ``mixing factor'' $({\alpha_i}/{\alpha_m})^2=a_m/a_i$.
In this case and in vacuum it is straightforward to obtain the usual wave equations for the electric (and similarly for the magnetic) field as
\begin{equation}
\frac{\partial^2(\alpha_i^2 \bar{E}^i)}{\partial t^2} = \left( \frac{\partial^2}{(\partial r^1)^2} + \frac{\partial^2}{(\partial r^2)^2} + \frac{\partial^2}{(\partial r^3)^2} \right) (\alpha_i^2 \bar E^i) \;,
\label{wave0}
\end{equation}
the solution of which is a field decaying as $\bar{E}^i(t) \propto 1/a_l a_m$.
This would simply mean that an EM wave in an anisotropic expanding space-time propagates in each direction with an intensity inversely proportional to the product of the two scale factors in the perpendicular directions.
In fact one would expect intuitively this result as due to the stretching of the expanding space directions.
But the inclusion of the mixing term between the scale factors in different directions introduces an extra coupling between the three components of the EM fields.
Eq.(\ref{wave0}) then becomes:
\begin{eqnarray}
\frac{\partial^2 \left(\alpha_i^2 \bar{E}^i\right)}{\partial t^2}
&=& \left( \frac{\partial^2}{(\partial r^1)^2} + \frac{\partial^2}{(\partial r^2)^2} + \frac{\partial^2}{(\partial r^3)^2} \right) (\alpha_i^2 \bar E^i) \nonumber \\
&+& \frac{1}{\alpha_i^2} \frac{\partial (\alpha_i^2)}{\partial t} \frac{\partial (\alpha_i^2 \bar E^i)}{\partial t}  \nonumber \\
&-& \epsilon_{ilm} \frac{\alpha_i^2}{\alpha_m^2} \frac{\partial (\alpha_m^2)}{\partial t} 
\frac{\partial \bar B^m}{\partial r^l} \;.
\label{wave1alpha}
\end{eqnarray}

Thus, in the absence of free charges and currents, the equation for the evolution of the electric field takes the following form, for example in the direction $1$:
\begin{eqnarray}
\frac{\partial^2 \left(a_2 a_3 \bar{E}^1\right)}{\partial t^2}
&=& \left( \frac{\partial^2}{(\partial r^1)^2} + \frac{\partial^2}{(\partial r^2)^2} + \frac{\partial^2}{(\partial r^3)^2} \right) (a_2 a_3 \bar E^1) \nonumber \\
&+& \left( \frac{\dot a_2}{a_2} + \frac{\dot a_3}{a_3} \right) \frac{\partial (a_2 a_3 \bar E^1)}{\partial t} \\
&-& \frac{1}{a_1} \left[ a_3 \frac{\partial (a_1 a_2)}{\partial t} \frac{\partial \bar B^3}{\partial r^2}
- a_2 \frac{\partial (a_1 a_3)}{\partial t} \frac{\partial \bar B^2}{\partial r^3} \right] \;. \nonumber
\label{wave1}
\end{eqnarray}
Comparing this equation with Eq.(\ref{wave0}), we see that the second and third terms in the right hand side of Eq.(\ref{wave1}) are due to the presence of the mixing factors in Eqs.(\ref{maxanis2}).
To solve this system of six coupled equations (one for each component of $\bar{E}$ and $\bar{B}$) is not trivial and is beyond the scope of this paper.

Nevertheless, one can see what happens in the particular case in which $a_2=a_3=a$.
In this case Eq.(\ref{wave1}) becomes:
\begin{eqnarray}
\frac{\partial^2 \left(a^2 \bar{E}^1\right)}{\partial t^2}
&=& \left( \frac{\partial^2}{(\partial r^1)^2} + \frac{\partial^2}{(\partial r^2)^2} + \frac{\partial^2}{(\partial r^3)^2} \right) \left(a^2 \bar{E}^1\right) \nonumber \\
&-& \left( \frac{\dot a_1}{a_1} - \frac{\dot a}{a} \right) \frac{\partial (a^2 \bar E^1)}{\partial t} \;.
\label{wave2}
\end{eqnarray}
By comparison with Eq.(\ref{wave0}), it is more evident the effect of the presence of the mixing factor.
It produces the second term in the right hand side, that introduces, via a mixing between the different components, a correction to the amplitude of the EM wave, that is dependent on the degree of anisotropy between one direction and the other two.
Equations like (\ref{wave2}) are well studied since they describe the propagation of EM waves in a conducting medium, and the solution is a dissipative-like wave with amplitude
scaling as $e^{-\gamma r}$, where $\gamma \simeq 1/2({\dot a_1}/a_1 - {\dot a}/a)$ over all the EM spectrum and $r$ is the direction of propagation.
This means that the amplitude of the field in one direction can increase or decrease, according to the fact that the Hubble expansion in that direction is smaller or larger than in the other directions.
Considering the conservation of the flux, the opposite would happen for the other components.
Thus the effect of the mixing factor in Eqs.(\ref{maxanis2}) is to transfer the fields between the different directions, with a dependence on the anisotropy opposite to what is described by Eq.(\ref{wave0}).
Clearly both of these effects have to be considered for a complete description of the behaviour of the EM wave, which in this case propagates with amplitude $\bar{\cal E}^1$ given by:
\begin{equation}
\bar{\cal E}^1 \simeq \bar{E^1_0}\,a^{-2}\,e^{-\frac{1}{2}\left(\frac{\dot a_1}{a_1} - \frac{\dot a}{a}\right) r} \;,
\label{sol1}
\end{equation}
where $\bar{E^1_0}$ is the initial amplitude.

A complete comprehension of Eqs.(\ref{maxanis2}) and their quantitative effects on the propagation of EM waves will require dedicated studies, but here I can qualitatively make some suggestions.

From the above simplified analysis it is evident that an anisotropic expansion of space-time induces a change of the amplitudes of the components of the EM fields, with the following expected consequences on the radiation emitted by a source:
\begin{itemize}
\item dependence of the amplitude on the space direction;
\item polarization of unpolarized light and rotation of polarization for linearly polarized light.
\end{itemize}
There is indeed also the possibility to explore the occurrence of refraction-like phenomena.

Considering the expected anisotropies in space-time expansion, as reviewed in the introduction, one can predict some astrophysical consequences.
The regions with anisotropic expansions are expected on large scales. 
Integrating the effect along the path of the light towards us, an observer sees the result of a collection of uncorrelated domains with different anisotropic properties. 
If the domains are not too small compared to the scale of the observable Universe, or better to the distance from the source to the observer, the integrated effect is not negligible.
One expects that the strength of this polarizing effect is inversely proportional to the distance, since, adding more domains with randomly distributed polarizing properties along the travel path of photons, essentially randomize the final polarization.
For example, the photons from the CMB, which travelled over more than ten Gpc before reaching our telescopes, have gone through more domains than photons emitted by quasars, travelling through a few Gpc, and thus the polarizing effect on the CMB is comparatively more suppressed.

As the strenght of this effect is very dependent on the degree of anisotropy of the expansion and on the space and time distribution of anisotropies, any estimate of it requires a careful modeling of the environment in which the EM waves propagate, and thus presents a large variability.
Anyway, just to give an idea of the numbers involved in the phenomenon, we may simply use Eq.(\ref{sol1}) to give a rough estimates in a particular situation. 
Considering an anisotropy in the Hubble rates of order 1-10\% over a distance of order 100 Mpc, an EM wave traveling through such a region would experience a relative change of the EM field in one direction of order 0.1-1\%.
This change would imply an extra linear polarization of the same amount.
Interestingly, this is similar to the order of magnitude of the extra polarization needed in one direction in order to explain the Hutsemekers effect.

In conclusion, whatever the anisotropic expansions are in the history of the Universe, I may predict the following effects:
\begin{itemize}
\vspace{-1mm}
\item polarization signals at scales of superclusters (some hundreds Mpc);
\vspace{-1mm}
\item polarization and rotation of polarization angles of light emitted by distant astrophysical sources, that could explain the ``Hutsemekers effect'' \cite{Hutsemekers:1998,Hutsemekers:2000fv,Hutsemekers:2005iz};
\vspace{-1mm}
\item possible polarization signals on the CMB, even if they should be suppressed;
\vspace{-1mm}
\item EM fields scaling differently in different directions from the source, meaning that what we measure is not necessarily representative of the intrinsic properties of the source, with possible connections with distance measurements.
\end{itemize}

The above predictions constitute a new observational tool to study the Universe at large scales, especially if combined with $H_0$ measurements, SN observations and gravitational lensing, and considering the polarization informations expected from Planck.

\vspace{4mm}

\begin{acknowledgments}
I am grateful to Jean-R\'en\'e Cudell, Yves De Rop, Damien Hutsemekers, Alexandre Payez and Ica Stancu for useful discussions on this and related subjects, and the IFPA department for inspiring hospitality.
I also acknowledge the financial support of the Belgian Science Policy.
\end{acknowledgments}

\bibliography{ED-anis-exp}

\end{document}